\documentstyle[12pt,epsfig]{article}
\let\section=\subsection     \let\subsection=\subsubsection                
\newcommand{\Psilrtf}{\Psi_{\lambda}(r,\vartheta,\varphi)}
\newcommand{\Nln}{{\cal N}_{ln} }
\newcommand{\jlkr}{j_l(k_{nl}r) }
\newcommand{\cYlmn}{{\cal Y}_{l}^{m}(\vartheta,\varphi) }

\newcommand{\CllV}{C_{\lambda \lambda'}^{(V)} }
\newcommand{\Cllt}{C_{\lambda \lambda'}^{(\tau)} }

\newcommand{\CllW}{C_{\lambda \lambda'}^{(W)} }
\newcommand{\CllS}{C_{\lambda \lambda'}^{(S)} }
\newcommand{\CllC}{C_{\lambda \lambda'}^{(C)} }
\newcommand{\vgrad}{\vec{\nabla}}

\begin{document}
\begin{center}
   {\large \bf INSTABILITIES}\\[2mm]
   {\large \bf OF A HOT EXPANDED NUCLEAR DROPLET}\\[5mm]
    \underline{P.~ROZMEJ}$\,^{a,b,}$\footnote{Talk given at XXVII
    International Workshop on Gross Properties of Nuclei and Nuclear
    Excitations, "MULTIFRAGMENTATION", Hirschegg, January 17--23, 1999},
    W.~N\"ORENBERG$\,^{a,c}$ and G.~PAPP$\,^{a,d}$  \\[5mm]
   {\small \it  $^a\,$Gesellschaft f\"ur Schwerionenforschung (GSI),
   D-64220 Darmstadt, Germany \\
   $^b\,$Instytut Fizyki, Uniwersytet MCS, 20-031 Lublin, Poland\\ 
   $^c\,$Institut f\"ur Kernphysik, Technische Universit\"at, Darmstadt, Germany\\
   $^d\,$Institut f\"ur Theoretische Physik, Universit\"at Heidelberg, Germany\\[8mm]}
\end{center}

\begin{abstract}\noindent
   The stability of hot expanded nuclear droplets against small bulk and surface oscillations is examined and possible consequences for multifragmentation are discussed.   
\end{abstract}

\section{Introduction}
\label{intro}

Spinodal instabilities were suggested by Bertsch, Siemens and Cugnon~\cite{bertsch-cugnon} already 15 years ago as a possible mechanism leading to multifragmentation of hot expanding nuclear matter. 

In heavy-ion collisions we expect that hot nuclear droplets are formed, which subsequently            
expand leading to low densities in the interior. Below certain values of the density, bulk and surface instabilities may occur and lead to multifragmentation.


In this contribution we report on a study of bulk and surface instabilities of spherical nuclear droplets as function of temperature and density.

\section{Bulk instabilities}
\label{bulk}

We consider the normal modes of a nuclear droplet. Let us define a complete orthonormal set of functions $\Psilrtf = \Nln \, \jlkr\, \cYlmn $, where $\cYlmn$ are modified (real) spherical harmonics,  $j_{l}(kr)$ are spherical Bessel functions and $\Nln$ are normalization constants. As we aim to consider the distortions of a spherical droplet with radius $R_0$, we impose the condition $j_{l}(k_{nl}R_0)=0$. We consider the distortions defined by the irrotational displacement field
\begin{equation}\label{gf7}
\vec{s}(\vec{r},t) = \vgrad \, \sum_{\lambda}\,
q_{\lambda}(t)\,\Psi_{\lambda}(\vec{r}) \equiv \vgrad \, w(\vec{r},t)\; 
\end{equation}
from the general form of the dispacement potential $w(\vec{r},t) = \sum_{\lambda}\,
q_{\lambda}(t)\,\Psi_{\lambda}(\vec{r}) $ . Note that the surface is not kept fixed
with this definition of the displacement field. The density varies according to the continuity equation
\begin{equation}\label{gf10}
\frac{\partial \varrho(\vec{r},t)}{\partial t} +
\mbox{div} [\varrho(\vec{r},t)\,\vec{v}(\vec{r},t)] = 0\; ,
\end{equation}
which guarantees exact conservation of mass, and -- together with the condition $\Psi_{\lambda}(R_0,\Omega)=0$ -- also of the center of mass.

In harmonic approximation small oscillations around equilibrium 
are determined by the set 
\begin{equation}\label{gf12}
 \sum_{\lambda} \, B_{\lambda \lambda'} \,\ddot{q}_{\lambda'} +
 \sum_{\lambda} \, C_{\lambda \lambda'} \,{q}_{\lambda'} =0 \; 
\end{equation}
of coupled equations. The eigenmodes are obtained from diagonalizing 
\begin{equation}\label{gf13}
(C_{\lambda \lambda'}-\omega^2\,B_{\lambda \lambda'})
{q}_{\lambda'} =0 \; .
\end{equation}
For ${\omega}^2>0$  the corresponding mode is stable (${q}_{\lambda} \sim \sin(\omega t)$). Exponential instability (${q}_{\lambda} \sim \exp(\gamma t)$) occurs for ${\omega}^2= -\gamma^2<0$.
\begin{table}
\begin{center}
\caption{Parameters of SkM$^*$ and SIII Skyrme forces}
\label{tab:1}       
\begin{tabular}{lccccccc}\hline\noalign{\smallskip}
& $t_0$ & $t_3$ & $t_1$ & $t_2$ & $x_0$ & $x_3$ & $\alpha$ \\ 
\noalign{\smallskip}\hline\noalign{\smallskip}
SkM$^*$ & -2645 & 15595 & 
410 & -135 & 0.09 & 0 & 1/6 \\ 
SIII & -1128.75 & 14000 & 395 & -95  & 0.45 & 1 & 1 \\ 
\noalign{\smallskip}\hline
\end{tabular}
\end{center}
\vspace*{-6mm}
\end{table}

Analytic expressions for the mass $(B_{\lambda \lambda'})$ and stiffness $(C_{\lambda \lambda'})$ tensors are derived from the velocity field $\dot{\vec{s}}$ and the Skyrme energy-density functional~\cite{brack}, respectively. The mass tensor is diagonal. The stiffness tensor $C$  is the sum of the contributions $C^{V}$,  $C^{\tau}$, $C^{W}$, $C^{S}$ and $C^{C}$ resulting, respectively, from the volume, intrinsic kinetic energy, Weizs\"acker, surface and Coulomb terms in the energy density. The analytic expressions of the tensors are given in~\cite{NPR}. Modes belonging to different $l,m$ (multipoles and their components) are decoupled. The only couplings left are those corresponding to different numbers $n$ of nodes in the displacement field for the same multipolarity. The eigenvalues are obtained by numerical diagonalization of the matrix $B^{-1}\,C$. The calculations have been performed for two Skyrme forces, i.e. SkM$^*$ and SIII implying, respectively, a soft and a stiff equation of state (EOS). The parameters of both forces are listed in Table~\ref{tab:1}. The contribution from the intrinsic kinetic energy has been calculated in the adiabatic limit (constant entropy with isotropic momentum distribution). 
 
\begin{center}
\begin{minipage}{13cm}
\centerline{\hspace*{3mm}{\epsfysize=75mm \epsfbox{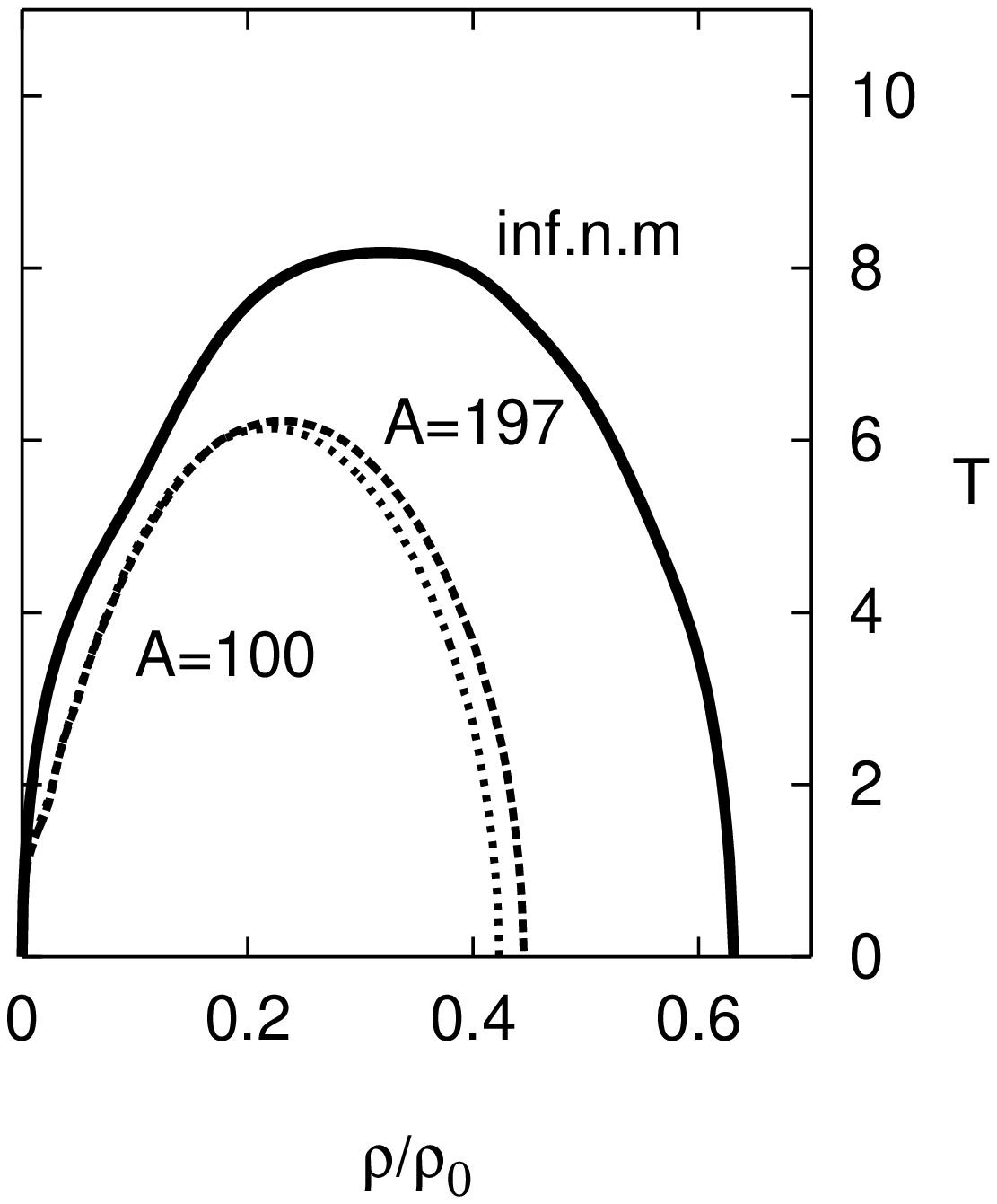}}
\hfill{\epsfysize=75mm \epsfbox{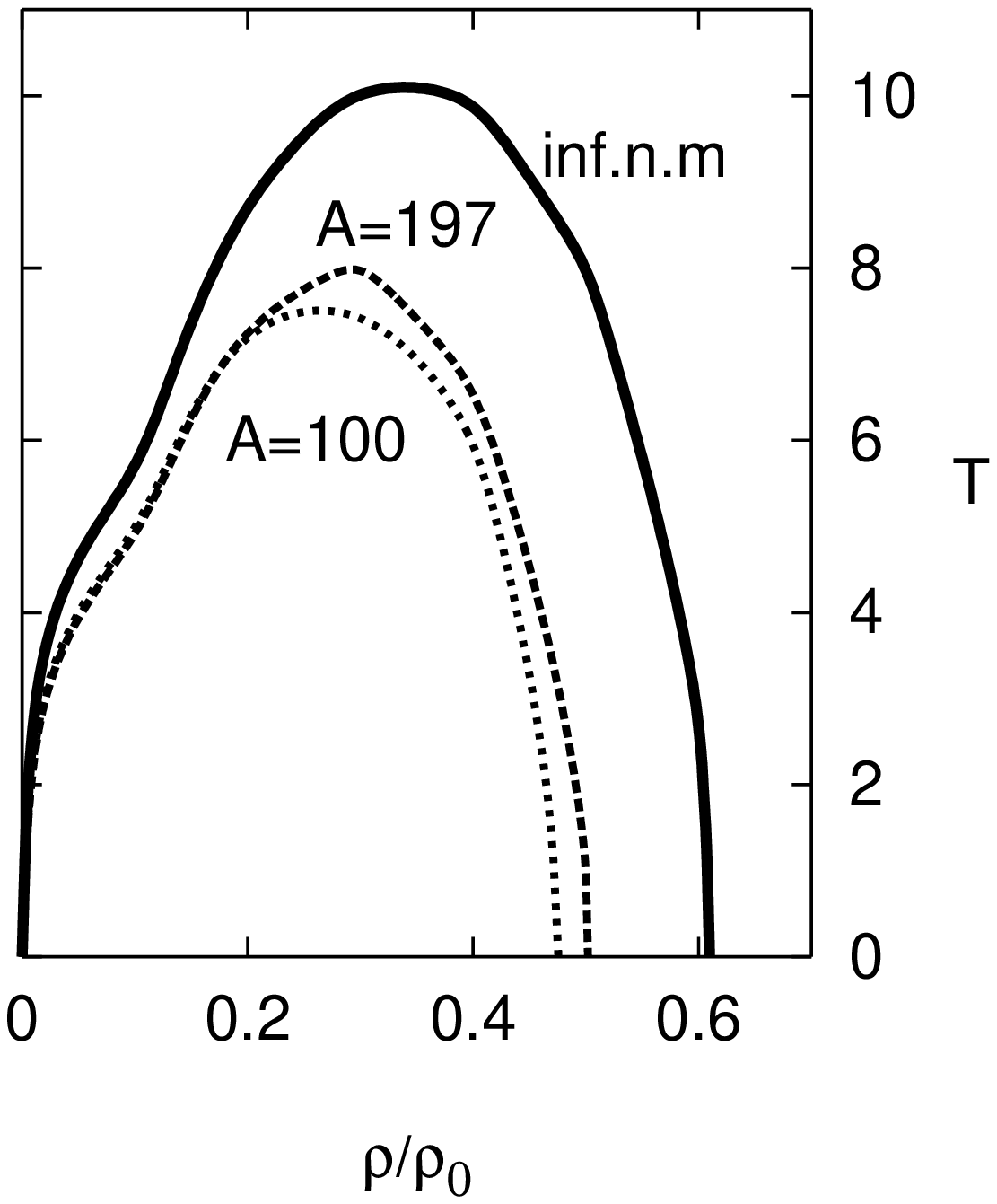}}}
\baselineskip=12pt
{\begin{small} Fig.~1. Borders of spinodal instabilities for infinite symmetric nuclear matter (heavy solid line), for a gold nucleus (A=197, dashed line) and for a smaller nucleus (N=Z=50, dotted line) as calculated for the soft (left) and the stiff (right) EOS, respectively. Temperatures $T$ are given in MeV. \end{small}}
\end{minipage}
\end{center}

From now on, for all figures, $\rho/\rho_0$ denotes the ratio of the actual density to normal nuclear density ($\rho_0 = 0.16$ fm$^{-3}$). Fig.~1 displays the areas of spinodal instability for the density modes with $l=2$ for three cases, i.e. infinite symmetric nuclear matter, a gold nucleus (Z=79, A=197), and a symmetric, roughly two times smaller fragment (Z=N=50) calculated with SkM$^*$ (left) and SIII (right) forces, respectively. For infinite nuclear matter the area of spinodal instability is substantially larger than those for finite systems. The difference between the spinodal lines of the two finite systems is small. The areas of spinodal instability for the higher multipoles $l=3,4,\ldots$ are further reduced. The results for infinite symmetric nuclear matter have been obtained by neglecting surface, Coulomb and Weizs\"acker terms ($\CllS= \CllC= \CllW=0$) and taking the limit ($R\rightarrow \infty$) in the volume and kinetic-energy terms ($\CllV$, $\Cllt$).

Fig.~2 shows quantitatively the importance of different contributions to the lowest eigenfrequency as function of the density at a typical temperature of 4~MeV. Due to their small values, the surface and Coulomb terms have practically no influence on the stability of the nuclear droplet. There is a delicate balance between the contribution from the kinetic energy term $\Cllt$ and the volume term  $\CllV$, such that the role of the Weizs\"acker term $\CllW$ becomes crucial. This term grows substantially with density and thus is important for the increasing stability with increasing density. 

\begin{center}
\begin{minipage}{13cm}
\centerline{{\epsfysize=80mm \epsfbox{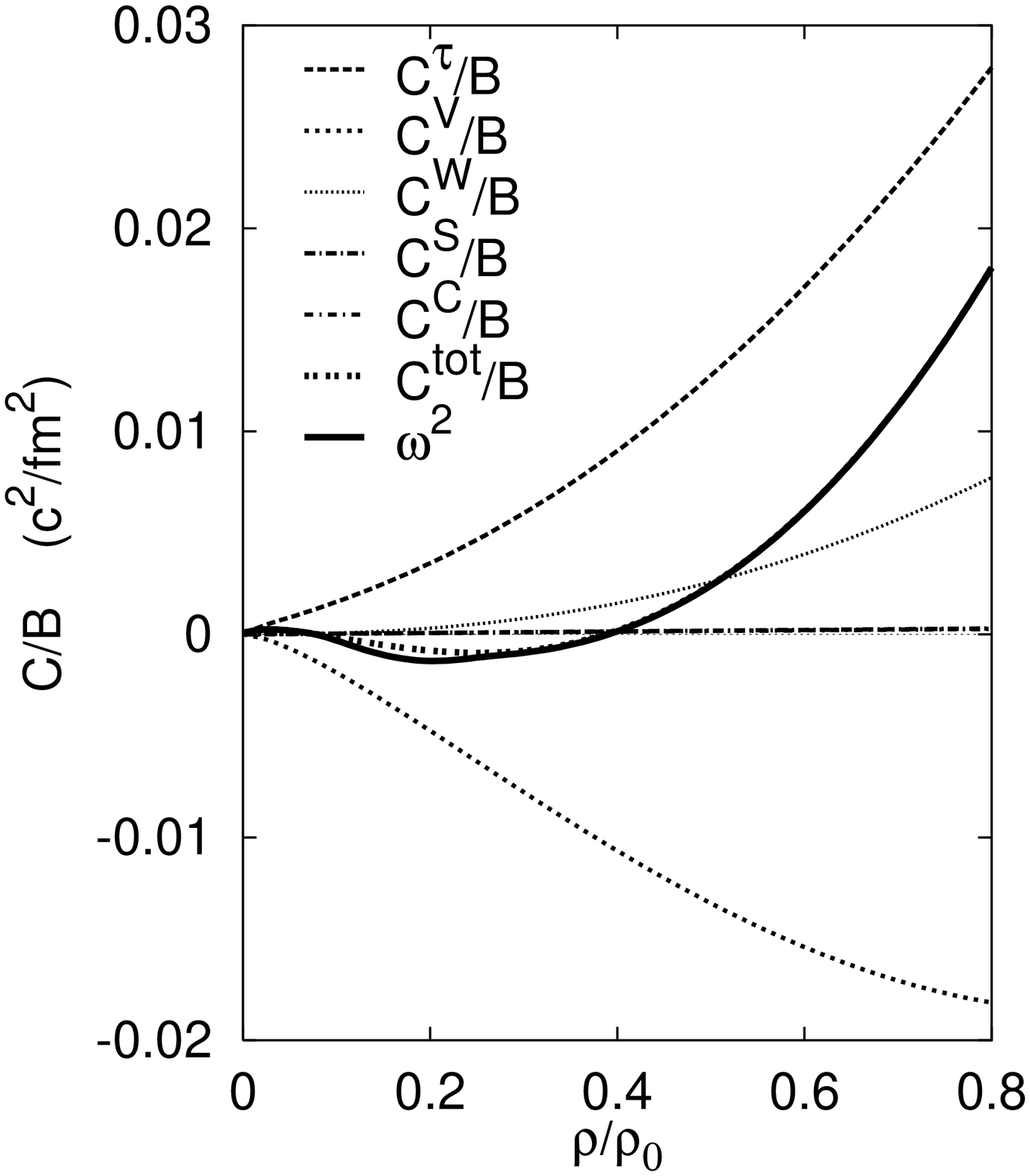}}
\hspace*{-3mm}{\epsfysize=80mm \epsfbox{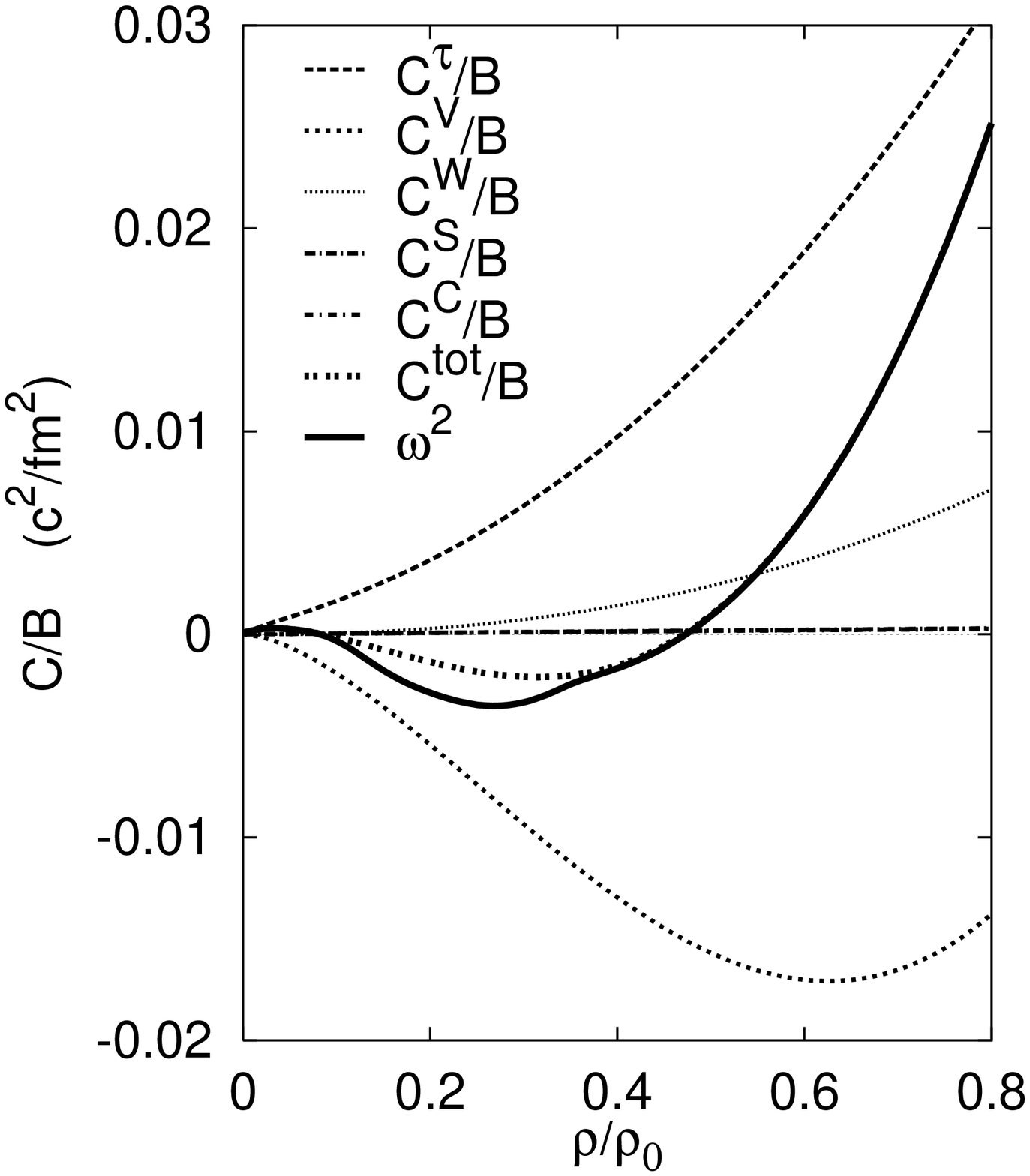}}}
\vspace*{5mm}
\baselineskip=12pt
{\begin{small} Fig.~2. Contributions of different terms to the lowest diagonal element of the $B^{-1}C$ matrix at $T=4$ MeV for a gold nucleus and soft (left) and stiff (right) EOS. Additionally the lowest eigenvalue $\omega^2$ is displayed by the solid line. The difference of this eigenvalue from $C^{tot}/B$ is due to couplings to higher-n modes. Note that contributions from Coulomb and surface terms are very small. \end{small}}
\end{minipage}
\end{center}

The crucial quantity in the multifragmentation process is the instability growth rate $\gamma$ (for $\omega^2<0$, $q_\lambda \sim \exp(\gamma t)$) or the characteristic growth time $\tau=1/\gamma$ for a particular mode. Multifragmentation, initiated by such instabilities, can occur only if these characteristic times are short enough compared to the characteristic evolution time of the system. In Fig.~3 (left) we present the shortest characteristic times as function of density and temperature for the bulk instabilities of the gold system calculated with the soft EOS. 

\section{Surface instabilities}
\label{surface}

For early stages of the expansion, where densities and temperatures are still high, no bulk instabilities exist. There instead, e.g. for initial temperatures  higher than 8 MeV, the surface vibrations are found unstable. However, the characteristic times for these
instabilities are about an order of magnitude larger than those for bulk oscillations in accordance with~\cite{ngo}. We follow here the standard Bohr and Mottelson theory \cite{B&M}.  
Fig.~3 (right) presents the smallest characteristic times for these surface instabilities. Although our basis is complete and -- in principle -- a linear combination of the collective displacement fields $\Psi_\lambda$ can describe an arbitrary collective motion, including surface vibrations of incompressible matter, such surface motion requires a very large number of terms due to slow convergence. Therefore, we calculated the surface instabilities separately. As we see from Fig.~3 bulk instabilities are dominant at small densities and temperatures, while for large enough temperatures and densities only surface modes are unstable.\\[-5mm]
\begin{minipage}{15.5cm}
\hspace*{-1.0cm}
\epsfysize=125mm \epsfbox{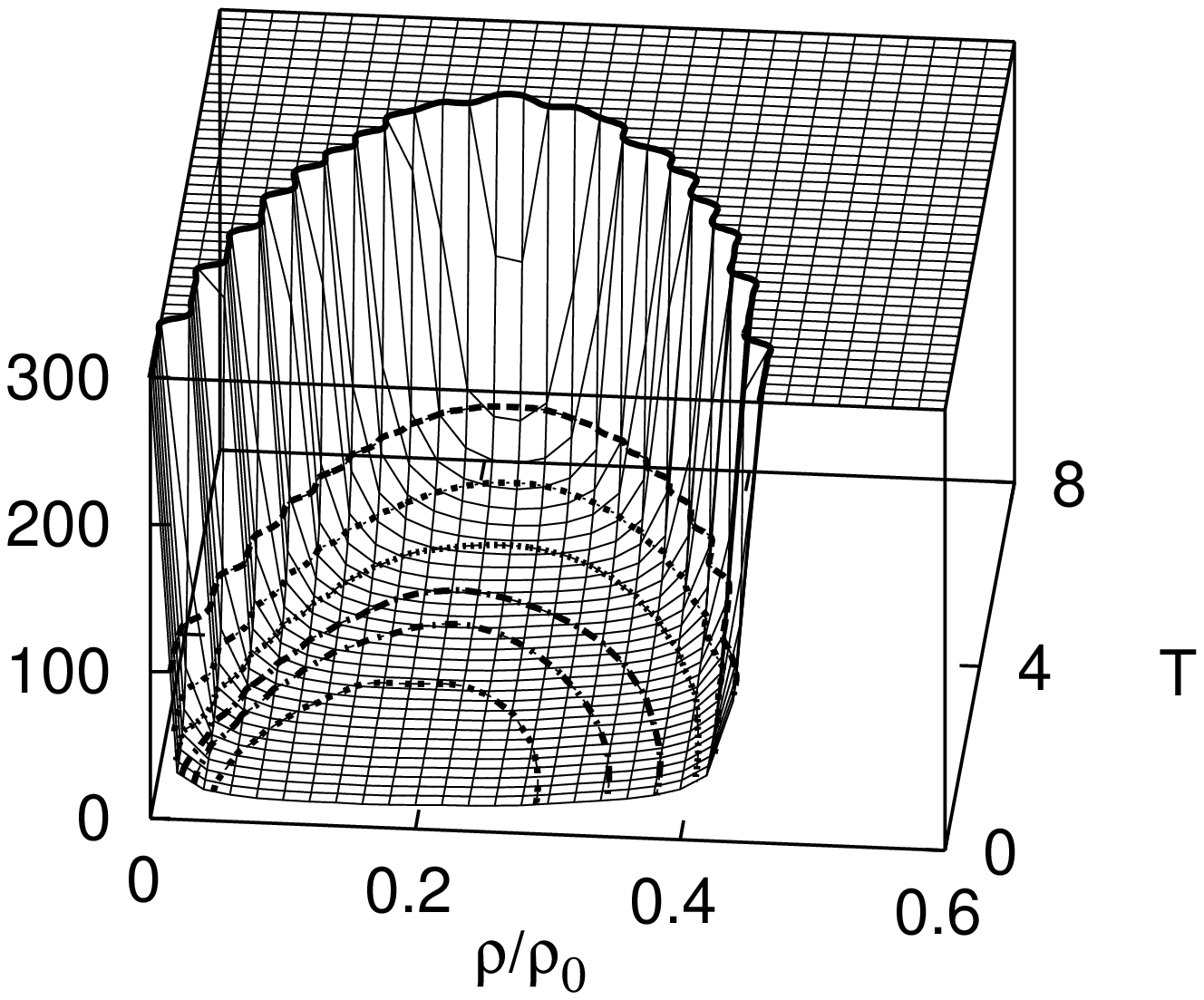}
\vspace*{-125mm}\hspace*{-1.7cm}
\epsfysize=125mm \epsfbox{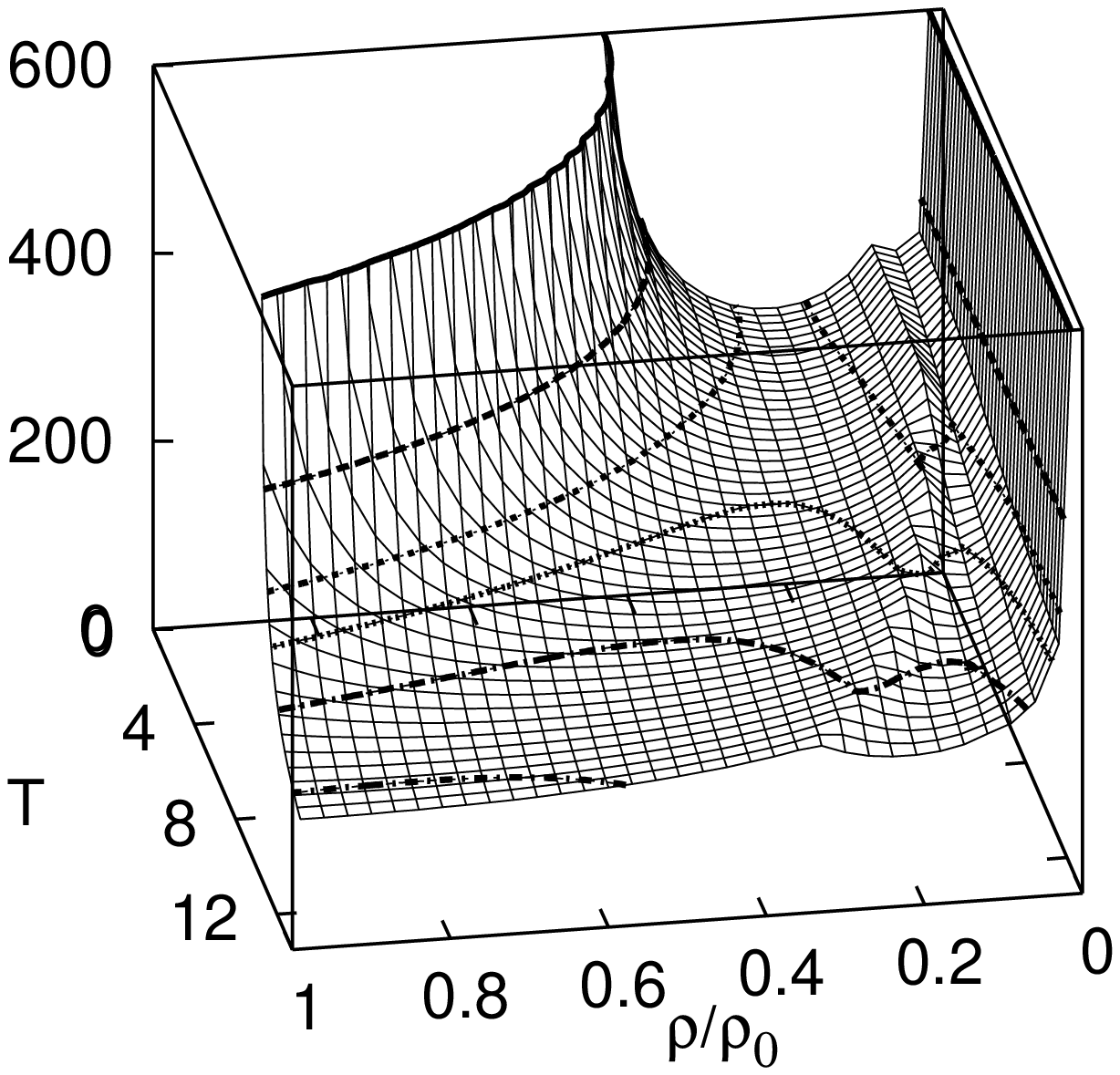}
\end{minipage}
\begin{center}
\vspace{6.7cm}
\begin{minipage}{13cm}
\baselineskip=12pt
{\begin{small} Fig.~3. Minimal characteristic times for bulk instabilities of a gold nucleus calculated with a soft EOS (left 3d-plot). All modes with $l=2,3,4,5$ and $n=1,\ldots6$ are taken into account. The contour lines correspond to the values 20, 25, 30, 40, 60 and 100 fm/c. The temperature $T$ is given in MeV.  The right 3d-plot shows the same (rotated by $\pi$ around the vertical axis) for surface instabilities. Here, the contour lines correspond to the values 150, 200, 250, 300, 400 and 600 fm/c. Note that in both diagrams the system is stable in the regions outside the holes.
\end{small}}
\end{minipage}
\end{center}

\section{Consequences for multifragmentation}
\label{consequences}

The expansion of hot nuclei has been studied in refs.~\cite{papp1-papp2-papp3} for soft and stiff equations of state and compared with experimental data \cite{herrmann-jeong-hsi-pochodzalla-eos}. For the soft EOS the expansion trajectories in the $(T,\rho)$ plane reach turning points around $T\approx 4$ MeV and $\rho/\rho_0\approx (0.25-0.45)$. Around these turning points the collective motion is very slow, and hence fast enough unstable modes can develop and initialize multifragmentation.  Although finite size effects substantially reduce the area of spinodal instabilities with respect to that of infinite nuclear matter, it is clear from Fig.~1 that the turning points essentially remain in the instability regime. From Fig.~3 (left) we see that characteristic times for density instabilities around this turning points are on the order of 25--40~fm/c. Such times seem to be short enough to cause an irreversible decay of the system towards multifragmentation. Usually several modes become unstable in this region, with close characteristic times, such that the production of many different fragments is possible. Our study of surface modes show that initially deformed nuclear  droplets can hardly become spherical, because the characteristic times for shape restoration is large compared to the expansion time. For initial temperatures larger than 8 MeV such initial deformation will tend even to grow.


%
\end{document}